\documentstyle[12pt]{article}
\title{Can the Unruh--DeWitt detector extract energy from the vacuum?}
\author{Hrvoje Nikoli\'c  \\
Theoretical Physics Division, Rudjer Bo\v{s}kovi\'{c} Institute, \\
P.O.B. 180, HR-10002 Zagreb, Croatia \\
{\normalsize hrvoje@faust.irb.hr} \\
}
\begin{document}
\maketitle
\begin{abstract}
The Unruh effect can be correctly treated only by using the Minkowski 
quantization and a model of a ``particle" detector, not by using the 
Rindler quantization. 
The energy produced by a detector accelerated only for a short time 
can be much larger than the energy needed to change
the velocity of the detector.
Although the measuring process lasts 
an infinite time, the production of the energy can be qualitatively 
explained by a time-energy uncertainty relation.
\end{abstract}  
\vspace*{0.5cm}

\section{Introduction}

A uniformly accelerated ``particle" detector  
coupled to a quantum field behaves as if it were  
immersed in a thermal bath with a temperature proportional to 
the acceleration \cite{unruh,dewitt}. Such a detector can 
spontaneously jump to a higher quantum level and produce 
a Minkowski particle. It is often argued that the energy needed 
for these two processes comes from the agency that accelerates 
the detector \cite{bd,padm1}. However, a detector can respond 
even without acceleration if the detection lasts a finite time. 
This effect is naturally interpreted as a consequence of  
time-energy uncertainty relations \cite{sva,padm2}. The aim of 
this letter is to demonstrate that even when the detection lasts 
an infinite time, the produced energy can be much larger than 
the energy needed to accelerate the detector. We do that by 
studying the response of a detector in a specific non-inertial 
trajectory, for which only a finite energy is needed for 
the acceleration. We find that this effect can also be qualitatively    
explained by a time-energy uncertainty relation.

There is an argument, based on the Rindler quantization, that the Unruh 
effect does not lead to a production of energy \cite{unruh2}. However, 
we show that the Rindler quantization is not a correct approach. First, 
the particle-model approach is not equivalent to the Rindler-quantization 
approach. Second, the event horizon, which plays the essential role 
in the Rindler quantization, cannot play any physical role for a local 
non-inertial observer. After discussing it in more detail in Sec. 2, we 
study the response of a pointlike detector in Sec. 3. The conclusions 
are drawn in Sec. 4, where the implications on the properties of the vacuum 
are also shortly discussed.           

\section{Inappropriateness of the Rindler quantization}

Let us first show that the ``particle"-detector approach to the Unruh 
effect based on the Minkowski quantization 
is not equivalent to the Rindler-quantization approach. For 
definiteness, we use the model of a monopole detector described in 
\cite{bd}. Assuming that the detector and the field are in the ground 
state $|0,E_0\rangle$ initially, the first order of perturbation 
theory gives the amplitude for the
transition to an excited state $|k,E\rangle$:
\begin{equation}\label{1}
A(k,\Delta E)=\bar{g} \int_{-\infty}^{\infty} d\tau \, e^{i\Delta E\,\tau}
\langle k|\phi(x(\tau))|0\rangle \; , 
\end{equation}
where $\bar{g}=ig\langle E|m(0)|E_0\rangle$, $g$ is a  
real dimensionless coupling constant, 
$m(\tau)$ is the monopole moment operator, $x(\tau)$ is the trajectory 
of the detector, $\Delta E=E-E_0$, and  
\begin{equation}\label{2}  
\langle k|\phi(x)|0\rangle =\frac{1}{\sqrt{(2\pi)^3 2\omega}} \, 
e^{i(\omega t -\bf{k}\cdot\bf{x})} \; .
\end{equation}
We compare the predictions that can be obtained from this model 
with the predictions that result from the Rindler quantization
\cite{unruh2,muller}. 

The 
Rindler-quantization approach predicts that the absorption      
of a Rindler particle by the accelerated atom will be seen 
by an inertial observer as an emission of a Minkowski particle
only if the atom has actually jumped to the
excited state. On the other hand, by putting $\Delta E=0$ in (\ref{1}), 
we see that the ``particle"-detector approach predicts an emission
of a Minkowski particle even if the transition to an excited state 
has not actually occurred. Nothing prevents the Minkowski particle 
produced in (\ref{1}) from being observed by the accelerated observer, 
contrary to the prediction of the Rindler quantization. For a uniform 
acceleration, the two approaches agree in the prediction of a thermal
distribution for $\Delta E$. However, even this partial agreement of the two
approaches does not generalize  
when the uniform acceleration is replaced by  
a more complicated motion \cite{padm3}.

The Rindler quantization is unitarily inequivalent to the Minkowski 
quantization. However, this fact, being an artefact of the infinite volume
\cite{gerlach}, is only a technical problem. A more serious problem 
is the fact that the Rindler quantization cannot be applied to the whole 
space-time, but only to the left and right wedges bounded by the event horizon 
\cite{fedotov}. Below we show that the event horizon does not correspond 
to any physical entity that could influence the properties of the fields seen 
by an accelerated observer, making the Rindler quantization physically 
meaningless. 

Let $x'$ be the Fermi coordinates of an observer at $\mbox{\bf{x}}'=\bf{0}$ 
moving arbitrarily in flat space-time.
If the observer does not rotate, the corresponding metric 
is given by $g'_{ij}=-\delta_{ij}$, $g'_{0i}=0$ and \cite{nels,nikolic1} 
\begin{equation}\label{1.1}
g'_{00}(t',\mbox{\bf{x}}')=(1+\mbox{\bf{a}}'(t')\cdot\mbox{\bf{x}}')^2 \; ,
\end{equation}
where $\mbox{\bf{a}}'$ is the proper acceleration. From (\ref{1.1}) we see that 
the Fermi coordinates of an accelerated observer possess a coordinate 
singularity at a certain $\bf{x}'$. However, in general, this coordinate 
singularity does not correspond to any physical boundary. Only 
$\bf{a}'(\infty)$, defining the event horizon, defines a physical 
boundary. However, in real life, acceleration never lasts infinitely long. 
And even if it does, it does not have any physical influence on a measuring 
procedure that lasts a finite time. Actually, the correct interpretation 
of the Fermi coordinates, and therefore also of the Rindler coordinates 
as their special case, is purely local \cite{nikolic1,nikolic2}, so 
they are not appropriate for quantization which requires a global approach
to describe the EPR-like correlations.   

\section{The response of the detector}

For a uniform acceleration, both the spent and the gained energy are
infinite, so it is not easy to compare them. A suspicion that the 
accelerating agency is not the source of the produced energy comes, 
for example, from the response of a detector in a uniform circular 
motion. The gained energy is infinite \cite{letaw,kim}, whereas the 
accelerating force does not raise the kinetic energy of the detector. 
Nevertheless, one could still argue that the force, acting 
during an infinite time, 
somehow gives energy to the quantum states of the detector. Therefore,  
we study a trajectory for which the force does not act infinitely long.     

We choose a simple trajectory with an instantaneous change of the velocity in 
the $z$-direction at the instant $\tau_0$, after and before which the detector  
moves inertially. Explicitly, $x=y=0$ and 
\begin{equation}\label{3}
z(\tau)=\left\{
\begin{array}{l}
0 \; , \;\;\; \tau\leq \tau_0 \; , \\
v\gamma(\tau -\tau_0) \; , \;\;\; \tau\geq \tau_0 \; , 
\end{array}\right.
\end{equation}  
\begin{equation}\label{4}   
t(\tau)=\left\{
\begin{array}{l} 
\tau \; , \;\;\; \tau\leq \tau_0 \; , \\
\gamma\tau +(1-\gamma)\tau_0 \; , \;\;\; \tau\geq \tau_0 \; ,
\end{array}\right.
\end{equation}  
where $\gamma=(1-v^2)^{-1/2}$. Calculating the divergent integrals of the 
type $\int_{\tau_0}^{\pm\infty}d\tau \, e^{i\Omega\tau}$ as 
\begin{equation}\label{5}
\lim_{\epsilon\rightarrow\pm 0^+ } \int_{\tau_0}^{\pm\infty}d\tau \,
e^{i\Omega\tau} e^{\mp\epsilon\tau} = \frac{i}{\Omega} \, e^{i\Omega\tau_0} \; ,
\end{equation}
from (\ref{1}), (\ref{2}), (\ref{3}) and (\ref{4}) we find
\begin{equation}\label{6}
A(k,\Delta E)=\frac{e^{i(\Delta E +\omega)\tau_0}}{i} \,  
\frac{\bar{g}}{\sqrt{(2\pi)^3 2\omega}} \, 
\left( \frac{1}{\Delta E +\omega} 
- \frac{1}{\Delta E +\gamma(\omega-k_z v)} \right) \; .
\end{equation}
For the case $v=0$, the amplitude vanishes, except for the trivial 
case $\Delta E =\omega=0$ (recall that $\Delta E$ and $\omega$ are 
non-negative). The physical quantity $|A|^2$ does not depend on $\tau_0$, 
just as we expect. Note also that if we replaced the instantaneous 
change of the velocity by a change that lasted a short but finite time, 
the result would not significantly change. We see that $\omega$, and 
therefore the gained energy $\omega+\gamma\Delta E$, can be arbitrarily 
large, although the kinetic energy $M(\gamma-1)$ spent for the change of the 
detector velocity is finite, provided that the mass $M$ of the detector is 
finite. Moreover, the averaged energy of the produced Minkowski 
particles 
\begin{equation}\label{7}
\langle \omega \rangle =\int d^3k \, \omega |A|^2 
\end{equation}
is linearly divergent. 

Although all further calculations can be performed using the exact expression 
(\ref{6}), we find it more instructive to employ the non-relativistic 
limit $v\ll 1$. In this limit, (\ref{6}) reduces to a simpler expression
\begin{equation}\label{8}
A(k,\Delta E)=ie^{i(\Delta E +\omega)\tau_0} \, 
\frac{\bar{g}}{\sqrt{(2\pi)^3 2\omega}} \, \frac{k_z v}{(\Delta E +\omega)^2}
\; .
\end{equation}
Note that (\ref{8}) implies that the average 3-momentum of the emitted 
Minkowski particles is zero, so one cannot object that the emission of the 
Minkowski particles affects the trajectory of the detector owing to the 
3-momentum conservation. 
Assuming that the mass of the scalar field is zero and using 
$d^3 k=d\varphi\, \sin \vartheta \, d\vartheta \, \omega^2 d\omega$ and  
$k_z=\omega\, \cos \vartheta$, we find 
\begin{equation}\label{9}
\langle \omega \rangle =\frac{\bar{g}^2 v^2}{3(2\pi)^2} 
\int_{0}^{\omega_{{\rm max}}} d\omega \, \frac{\omega^4}{(\Delta E
+\omega)^4} \; ,
\end{equation}
where the cut-off $\omega_{{\rm max}}$ is introduced only for the sake 
of regularization. Since the largest contribution to the divergent integral 
in (\ref{9}) comes from large $\omega$, we can take $\Delta E=0$, which 
leads to a simple expression    
\begin{equation}\label{10}
\langle \omega \rangle =\eta E_{{\rm spent}} \; ,
\end{equation}
where $E_{{\rm spent}}=Mv^2/2$ is the spent energy, while $\eta$ is the 
efficiency factor 
\begin{equation}\label{11}
\eta=\frac{\bar{g}^2}{6\pi^2}\, \frac{\omega_{{\rm max}}}{M} \; .
\end{equation}
Even if one assumes that ultraviolet divergences, typical for quantum 
field theory, require introduction of a finite cut-off $\omega_{{\rm
max}}$, nothing prevents (\ref{11}) from being larger than one.

If, as our results suggest, the energy of the accelerating agency is not the 
source of the produced energy $\omega+\Delta E$, then we must conclude that this 
disbalance of the energy is of quantum-mechanical origin. To support 
this conclusion, below we show that, although the measuring process lasts 
an infinite time, the disbalance of the energy can be qualitatively 
explained by a time-energy uncertainty relation. In our calculations we have 
assumed that the time $\tau_0$, at which the change of the velocity occurs, is 
known with certainty. On the other hand, if $\tau_0$ is completely 
unknown, then we must sum all the amplitudes (\ref{8}) with different 
$\tau_0$, which gives a vanishing total amplitude (except for the trivial 
case $\Delta E=\omega=0$). To explore an intermediate case of a finite 
uncertainty $\Delta\tau$ of the time at which the velocity changes, we introduce 
the averaged amplitude
\begin{eqnarray}\label{12}
\bar{A} & = & \frac{1}{\Delta\tau}
 \int_{\tau_0-\Delta\tau/2}^{\tau_0+\Delta\tau/2} d\tau_0 \, A \nonumber \\
& = & -e^{i(\Delta E +\omega)\tau_0}\, \frac{\bar{g}k_z v}{\sqrt{(2\pi)^3
 2\omega} \, (\Delta E +\omega)^2} \, \frac{\sin(\Delta E +\omega)\Delta\tau/2}
 {(\Delta E +\omega)\Delta\tau/2} \; . 
\end{eqnarray}
Now the averaged energy of the produced Minkowski particles is
\begin{eqnarray}\label{13}
\langle \omega \rangle & = & \int d^3k \, \omega |\bar{A}|^2 \nonumber \\
& = & \frac{\bar{g}^2 v^2}{3(2\pi)^2} 
 \int_{0}^{\infty} d\omega \, \frac{\omega^4}{(\Delta E 
 +\omega)^4} \, \frac{\sin^2 (\Delta E +\omega)\Delta\tau/2}
 {[(\Delta E +\omega)\Delta\tau/2]^2} \; .
\end{eqnarray} 
If one expands the sine in (\ref{13}) for small $\Delta\tau$ and 
retains only the lowest contribution, then one recovers (\ref{9}). 
However, (\ref{13}) with a non-zero $\Delta\tau$ is a finite quantity 
and is of the order
\begin{equation}\label{14}
\langle \omega \rangle \sim \frac{\bar{g}^2 v^2}{(\Delta\tau)^2 \Delta E} \; .
\end{equation}
This can also be written in the form (\ref{10}), with
\begin{equation}\label{15} 
\eta\sim\frac{\bar{g}^2}{(M\Delta\tau)(\Delta E\Delta\tau)} \; ,
\end{equation} 
revealing that the produced energy is larger when $\Delta\tau$ is smaller. 
Applying quantum kinematics to the motion of the detector, we estimate 
$(\Delta\tau)^{-1}\sim Mv^2/2=E_{{\rm spent}}$. Therefore, (\ref{15}) 
gives
\begin{equation}\label{16}
\eta\sim\bar{g}^2 v^2 \frac{E_{{\rm spent}}}{\Delta E} \; .
\end{equation}
This efficiency factor is finite. However, since for a typical case 
$E_{{\rm spent}}\gg \Delta E$, the efficiency factor can be much larger than 
one. Actually, from (\ref{8}) we see that the case $\Delta E =0$ is the most 
probable. From (\ref{16}) we see that this case leads to the largest
efficiency. The infrared divergence can be removed, for example, by taking 
the mass $m$ of the scalar particle to be non-zero, which leads to 
\begin{equation}\label{17}
\eta\sim\bar{g}^2 v^2 \frac{E_{{\rm spent}}}{m} \; .
\end{equation}   
Again, $\eta$ can be much larger than one. 

Equations (\ref{16}) and (\ref{17}) also suggest that $\eta$ is much 
larger for relativistic velocities $v$. This can also be explicitly 
shown by a similar calculation for the ultrarelativistic limit $\gamma\gg 1$ 
in (\ref{6}). 

\section{Conclusion}

Our analysis suggests that energy can be extracted from the vacuum. However, 
we stress that we have not found that the efficiency $\eta$ {\em must} be 
much larger than one. The coupling constant $\bar{g}$ or the velocity $v$ 
can be so tiny that $\eta$ is much smaller than one. 
Therefore, the fact that a large energy production has not yet been observed
is not in contradiction with our 
results. Our results suggest only that a large energy can be 
extracted {\em in principle}. This is not in contradiction with the 
conservation of energy if one accepts the picture of the vacuum as a state 
of infinite (or large, owing to a large cut-off) energy density, as, 
indeed, quantum field theory suggests. 

The only real problem with such a picture is to explain the smallness  
of the cosmological constant. A possible way out of this problem is to 
propose that only excited states, i.e. particles, contribute to 
$\langle T_{\mu\nu}\rangle$ in the semi-classical Einstein equation. 
The results of Sec. 2 and those of \cite{nikolic2} suggest that there exist 
preferred coordinates with respect to which fields should be quantized  
and, consequently, that the notion of a particle does not depend on the
observer, making such a proposal consistent. Another interesting 
possibility of resolving the cosmological constant problem is to 
adopt the DeBroglie--Bohm interpretation of quantum field theory 
(which also requires a preferred time coordinate \cite{holl}). Since the 
ground-state wave function is real, its energy is exactly canceled by the 
quantum potential \cite{holl2,squ}.

\section*{Acknowledgement}

This work was supported by the Ministry of Science and Technology of the
Republic of Croatia under Contract No. 00980102.

\end{document}